\documentclass[a4paper, amsfonts, amssymb, amsmath, reprint, showkeys, nofootinbib, twoside]{revtex4-1}
\usepackage[english]{babel}
\usepackage[utf8]{inputenc}
\usepackage[colorinlistoftodos, color=green!40, prependcaption]{todonotes}
\usepackage{amsthm}
\usepackage{mathtools}
\usepackage{physics}
\usepackage{xcolor}
\usepackage{graphicx}
\usepackage[left=23mm,right=13mm,top=35mm,columnsep=15pt]{geometry} 
\usepackage{adjustbox}
\usepackage{placeins}
\usepackage[T1]{fontenc}
\usepackage{lipsum}
\usepackage{csquotes}

\usepackage[pdftex, pdftitle={Article}, pdfauthor={Author}]{hyperref} 

\newcommand{\p}{\text{P}}

\newcommand{\om}{\omega}
\newcommand{\ep}{\varepsilon}

\newcommand{\vta}{\vartheta}
\usepackage{array}

\begin{document}
\title{Mid-infrared frequency combs and staggered spectral patterns in $\chi^{(2)}$ microresonators}

\author{N.~Amiune$^{1}$, Z.~Fan$^{2}$, V.V.~Pankratov$^{2}$, D.N.~Puzyrev$^{2}$, D.V.~Skryabin$^{2,\dagger}$, K.T.~Zawilski$^{3}$, P.G.~Schunemann$^{3}$, and I.~Breunig$^{1,4,*}$}
\affiliation{\footnotesize{
		\mbox{$^{1}$Laboratory for Optical Systems, Department of Microsystems Engineering - IMTEK, University of Freiburg, Georges-K\"ohler-Allee 102,}\\ 79110 Freiburg, Germany\\
		\mbox{$^{2}$Department of Physics, University of Bath, Bath BA2 7AY, England, United Kingdom}\\
		\mbox{$^{3}$BAE Systems Inc., MER15-1813, P.O. Box 868, Nashua, New Hampshire 03061-0868,USA}\\
		\mbox{$^{4}$Fraunhofer Institute for Physical Measurement Techniques IPM, Georges-K\"ohler-Allee 301, 79110 Freiburg, Germany}}\\
	$^\dagger$d.v.skryabin@bath.ac.uk\\
	$^*$optsys@ipm.fraunhofer.de 
	}

\date{\today} 

\begin{abstract}
The potential of frequency comb spectroscopy has aroused great interest in generating mid-infrared frequency combs in the integrated photonic setting. However, despite remarkable progress in microresonators and quantum cascade lasers, the availability of suitable mid-IR comb sources remains scarce. Here, we present a new approach for the generation of mid-IR microcombs through cascaded three-wave-mixing. By pumping a CdSiP$_2$ microresonator at 1.55~µm wavelength with a low power continuous wave laser, we generate $\chi^{(2)}$ frequency combs at 3.1~\textmu m wavelength, with a span of about 30~nm. We observe ordinary combs states with a line spacing of the free spectral range of the resonator, and combs where the sideband numbers around the pump and half-harmonic alternate, forming staggered patterns of spectral lines. Our scheme for mid-IR microcomb generation is compatible with integrated telecom lasers. Therefore, it has the potential to be used as a simple and fully integrated mid-IR comb source, relying on only one single material.
\end{abstract}

\maketitle

Frequency comb sources in the mid-infrared (mid-IR) wavelength range present themselves as a promising tool for spectroscopy, with several works already showing improved features with respect to state-of-the-art spectrometers\cite{Picque2019}. In particular, dual-comb spectrometry attracts currently most interest, as it provides fast and accurate measurements without any moving parts\cite{Coddington2016}.
In sight of these applications, there is also increasing interest in microresonator and (interband) quantum cascade laser (QCL) combs to achieve a miniaturized mid-IR frequency comb source in the fingerprint region, where most molecules exhibit many strong absorption lines. Such technologies may enable, in the long-term, the realization of comb-based spectrometers fully on a chip-scale\cite{Scalari2019,Gaeta2019}. 

At the present time, however, the availability of compact and broad mid-IR comb sources is still highly limited despite the progress in these fields. With microresonators, the direct generation of Kerr combs in the mid-IR is typically realized by using a bulk optical parametric oscillator as a pump source\cite{Wang2013,Griffith2015,Luke2015,Yu2016,Yu2018}, which is not suitable for a fully integrated device. Using a QCL as a pump source was also demonstrated\cite{Savchenkov2015} but with comb line spacings too large ($\approx$ 40~THz) for spectroscopy. Another possibility is converting near-infrared combs into the mid-infrared via difference frequency generation\cite{Bao2020,Bao2021} or optical parametric oscillation\cite{Herr2018}. Regarding QCL combs, mode-locking in the mid-IR has been demonstrated with high optical powers and appear as a promising candidate, but are still limited in bandwidth\cite{Hugi2012,Feng2018,Bagheri2018,Sterc2020,Sterc2021}.

\begin{figure}[htb]
	\centering{		\includegraphics[width=0.48\textwidth]{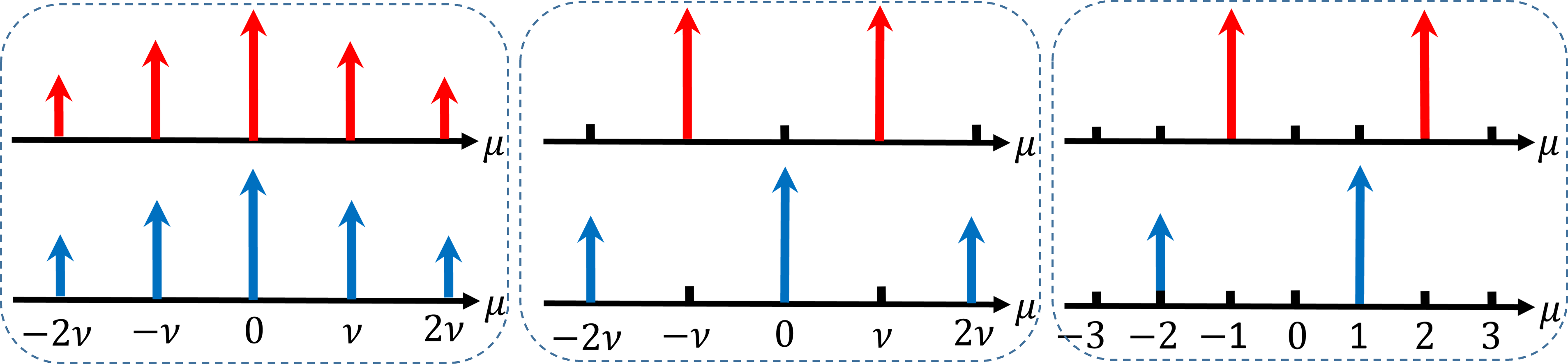}}
	\caption{The left column illustrates an ordinary comb with a spatial period $2\pi/\nu$, where $\nu$ is an integer, and $\mu$ is the relative mode number. The middle column shows a symmetric staggered comb with a spatial period $2\pi/2\nu$. The right column is an example of an asymmetric staggering corresponding to a spatial period $2\pi/3$. Red and blue colours mark the signal and pump combs, respectively. }
	\label{f0}
\end{figure}

Here, we introduce a new approach to generate frequency combs in the mid-infrared region by means of cascaded three-wave-mixing. By pumping a resonator with a $\chi^{(2)}$ nonlinearity, phase-matched for optical parametric oscillation (OPO) at degeneracy, a series of cascaded three-wave-mixing processes will lead to the generation of combs around the pump frequency and around half the pump frequency\cite{Ricciardi2020}. This scheme was first realized in bulk mirror cavities\cite{Ulvila2013,Ulvila2014,Mosca2018} and more recently in bulk\cite{Amiune2021} and chip-integrated\cite{Bruch2021} microresonators, as well as in waveguide cavities\cite{Wang2022}. In this work, we apply this concept with a CdSiP$_2$ mm-sized resonator in order to generate a frequency comb at 3.1~\textmu m by pumping at 1.55~\textmu m wavelength. This way, we generate 30~nm wide frequency combs, with mW pump powers and different repetition rates of 27.7~GHz and multiples. This scheme has the potential to be fully integrated, as it only requires a low power telecom continuous wave laser as a pump source and the integration of a single material. 
Furthermore, it has been recently predicted by us theoretically~\cite{opo}  that the  non-solitonic modelocked frequency combs, i.e., Turing pattern combs, 
in optical parametric oscillators (OPOs) can be spectrally staggered, see an illustration in the middle column Fig.~\ref{f0}. 
Here, we provide the experimental confirmation of the existence of such combs. Staggering is a unique feature of the OPO combs related to the fact that the central mode of the half-harmonic field is allowed to have zero power, which is forbidden in the sister case of the $\chi^{(2)}$ microcombs due to second-harmonic generation~\cite{Szabados20,shg}. The resulting comb pattern has an alternating sideband number between the pump and half-harmonic. We also report here combs with asymmetric staggering, sketched in the right column of Fig.~\ref{f0},
emerging when the pump couples to a resonance with an absolute odd longitudinal mode number instead of even.

\begin{figure*}[h]
	\centering{\includegraphics[width=0.9\textwidth]{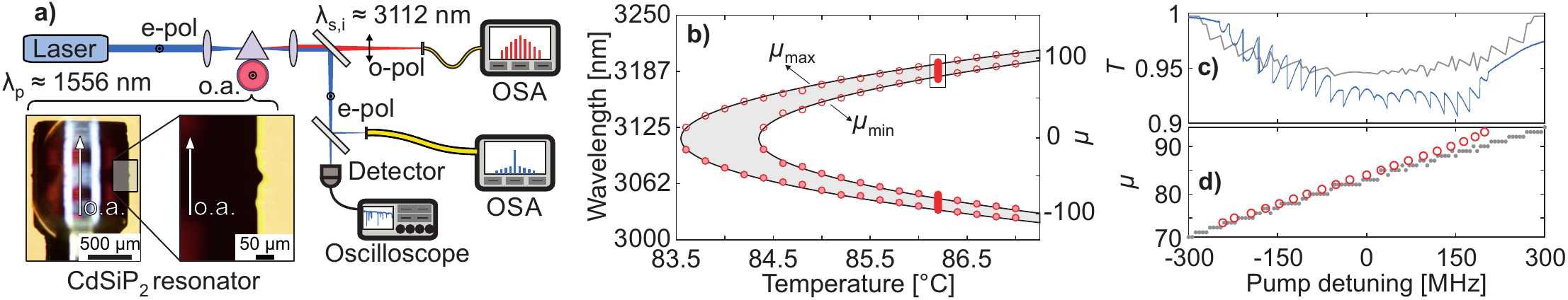}}
	\caption{\textbf{Experimental setup and OPO tuning to degeneracy.} a) Sketch of the experimental setup. OSA: Optical spectrum analyzer, o.a.: optic axis. b) OPO tuning curve towards degeneracy, where the output wavelengths depend on the relative pump detuning from resonance. The points correspond to the experimentally measured minimum and maximum wavelengths for a fixed temperature, while the curves are calculated. Fig. c) shows the pump transmission and d) the excited resonator mode as a function of pump detuning for the enclosed region in Fig. a). The blue line and red dots correspond to experimental measurements, while the gray lines and dots correspond to the numerical simulation.}
	\label{fig:setup}
\end{figure*}

A schematic of our experimental setup is shown in Fig.~\ref{fig:setup}a). We couple light from a continuous wave 1556~nm diode laser into a CdSiP$_2$ resonator with a radius $R = 560$~\textmu m, and we rely on its birefringence to fulfil the phase matching conditions. 
The optic axis is parallel to the symmetry axis of the resonator, and the pump light is extraordinarily polarized, while the down-converted signal is ordinarily polarized. For in-coupled pump powers below 100~µW we observe mid-IR light generation ranging from 2.3~µm to 5~µm depending on the transverse mode selected. By initially setting the temperature to $87^{\circ}$C we observe OPO operation close the point of degeneracy around 3112~nm, which agrees with our predicted phase-matching temperature considering fundamental transverse modes for all waves. We then proceed to tune the OPO output further to the point of degeneracy by reducing the temperature, and increasing the laser frequency to keep the pump light close to resonance. 

As the laser frequency is scanned across the resonance for a given temperature, the frequency of the output signal/idler waves can vary by hundreds of GHz. This occurs as the signal and idler waves oscillate in different longitudinal modes of the resonator depending on the pump detuning. The tuning curve  in Fig.~\ref{fig:setup}b) displays the measured minimum and maximum signal/idler wavelengths for different temperatures and in-coupled pump powers of 350~µW. The curves are fitted with the equation~\cite{opo},
\begin{equation}
    \mu_{\max,\min}^2 = - \frac{\ep_0}{D_{2s}} \mp \frac{\kappa_p}{\kappa_s} \frac{\gamma_s}{D_{2s}} \sqrt{\frac{W D_{1p}}{2\pi \kappa_p}}, ~~D_{2s}<0,
\label{eq1}
\end{equation}
where $\mu$ is the relative mode number counted from 
the resonator mode nearest to the pump frequency $\omega_p$, and $\ep_0=2\omega_{0s}-\omega_{0p}$ is the frequency mismatch 
parameter, see Supplemental Material (SM) for details. $D_{1p}/2\pi=27.06$~GHz is the pump repetition rate, $\kappa_p/2\pi=55$~MHz and $\kappa_s/2\pi=64$~MHz are the pump and signal linewidth, $D_{2s}/2\pi=-164$~kHz is the signal dispersion, $\gamma_s/2\pi=2$~GHz/$\sqrt{\text{Watt}}$ is the signal nonlinear coefficient, and $W=130$~\textmu W is the laser power matching the experimental data. Temperature dependence is introduced via $\ep_0(T) \approx \alpha (T-T_0)$, where $\alpha=464$~MHz/°C can be calculated from the resonant frequencies approximation, see SM. The value for $T_0$, corresponding to $\ep_0=0$, is fitted to the experimental data.

Values for signal and idler wavelengths inside the shaded area Fig.~\ref{fig:setup}b) can be selected in steps of one~FSR by thermally locking the pump laser to the resonance, as shown for the enclosed region at $T=86.2$~°C. For this temperature, Fig.~\ref{fig:setup}c) shows the pump transmission (blue is experiment and gray is numerical modelling).  
A series of dips corresponds to the excitation of different consecutive longitudinal modes of the resonator as the laser frequency is scanned across resonance. 
Fig.~\ref{fig:setup}d) shows a sequence of the measured (red circles) and numerically modelled (small black dots) resonator mode numbers
excited by  tuning the pump frequency. The experimentally determined mode numbers are calculated from the measured output wavelengths with $D_{1p}/2\pi$, and the detuning is aligned to the transmission dips. The numerical data reveal the existence of the locking intervals for every signal-idler pair, i.e., $\pm\mu$, and the respective ladder-like sequence of transitions from one interval of $\om_p$ to the next. The matching of the experimental and numerical data in Fig.~\ref{fig:setup} d) should be considered as the
experimental evidence of the theoretically predicted 
equivalence between the microresonator 
OPO tuning and the so-called Eckhaus instabilities, 
see Ref.~\cite{opo}.

\begin{figure*}[h]
	\centering{	\includegraphics[width=0.9\textwidth]{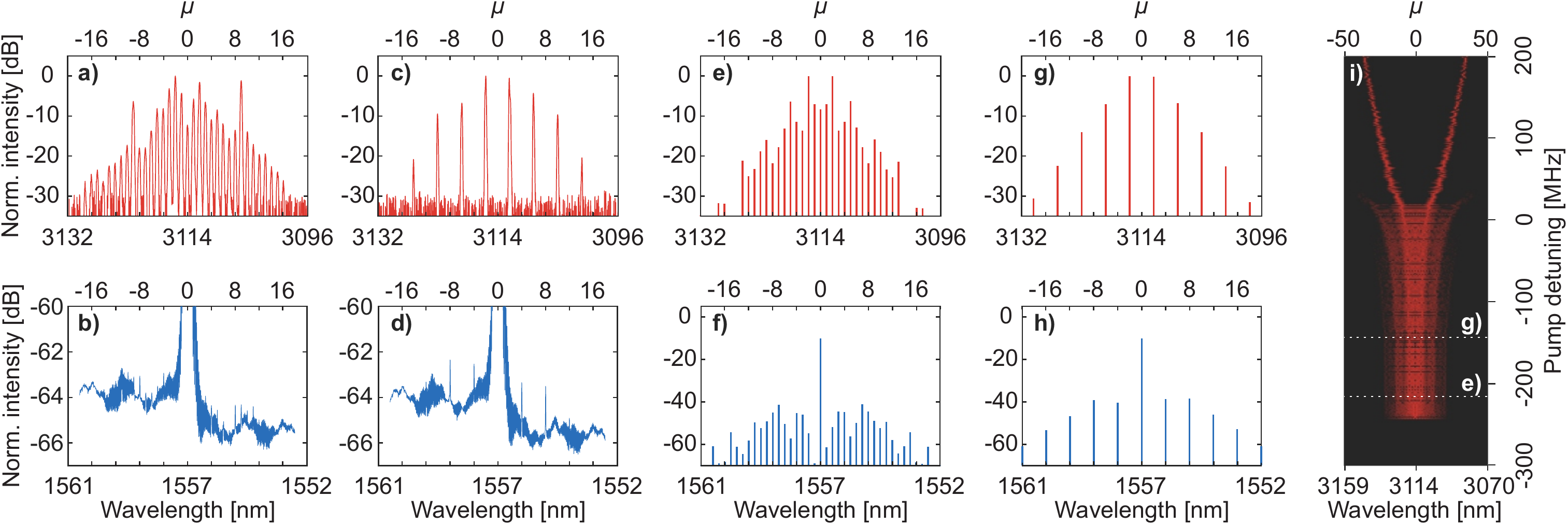}}
	\caption{\textbf{Ordinary and symmetrically staggered frequency combs for the pump coupled to the even mode number.} Frequency comb spectra around the pump (blue) and sub-harmonic signal (red) frequencies. In the experiment, the comb state changes from the  one FSR comb line spacing in a,b) to  the four FSR spacing in c,d) as the pump frequency is reduced across the resonance.  e-i) Numerically simulated spectra  for different pump detunings. e,f) and g,h) show the spectra for $- 216$ and $- 144$~MHz detunings, respectively. The background in i) is at -70~dB and $\ep_0=0$. c,d) and g,h) show the staggered combs.}
	\label{fig:evencomb}
\end{figure*}

In order to generate the frequency combs, we set the temperature and pump wavelength to operate in between the tips of the tuning curves from Fig.~\ref{fig:setup}b). Then, as we slowly reduce the pump frequency across the resonance (increasing detuning), we observe a transition from comb states to non-degenerate OPO with the signal and idler frequencies getting further apart. For an in-coupled pump power of 3~mW, we first observe a dense frequency comb around 3.1~\textmu m, shown in Fig.~\ref{fig:evencomb}a), with a width of about 30~nm. At the same time, we observe a weak comb around 1.55~\textmu m, shown in Fig.~\ref{fig:evencomb}b), approximately 60~dB below the pump power. Both the pump and signal comb lines are separated by one FSR and, therefore, provide an example of the ordinary comb structure, like the one illustrated in the left column of Fig.~\ref{f0}. 
Our numerical modelling has reproduced such combs, Figs.~\ref{fig:evencomb}e), f), and revealed that they correspond 
to a variety of complex wave forms which could be either stationary or weakly breathing.

By further reducing the laser frequency we observed a different type of comb, see Figs.~\ref{fig:evencomb}c),d) and g),h) for the experimental and numerical data, respectively. Now, the line spacing equals four FSRs for both pump and signal combs, but the signal sidebands are counted from $\mu=2$ and the pump ones form $\mu=0$. 
To the best of our knowledge, this is the first experimental evidence of the staggered OPO combs after their recent theoretical prediction~\cite{opo}. We shall note, that in this case, we are pumping in an even absolute mode number of the resonator, which is evident, e.g.,  from Fig.~\ref{fig:evencomb}c) where the spectral center of the practically symmetric comb occurs in the middle between the two lines.  
Our modelling shows that this state is a phase locked one. It belongs to the general category of the two-colour Turing pattern combs, see \cite{shg} and references therein. 

Conversion efficiency of the sidebands from the signal
comb (red spectra) back to the pump (blue spectra) 
is hampered by the large group velocity (repetition rate) difference, $(D_{1p}-D_{1s})/2\pi=0.51$GHz, which strongly dominates over dispersion and, therefore, makes all the three wave mixing processes, excluding the $\mu=0$ (pump)  to  $\mu=\pm 4$ (signal), to be significantly mismatched~\cite{opo,shg}. However, the reason for the lower sideband power observed experimentally compared to the simulations still requires further studies.

The near-IR (blue coloured) and mid-IR (red coloured) combs share the same repetition rate, which is a multiple of $27.7\pm 0.1$~GHz. This value lies closer to the linear FSR at 3.1~\textmu m ($D_{1s}/2\pi=27.57$~GHz), than to the pump FSR  ($D_{1p}/2\pi=27.06$~GHz). The selection of the nonlinear repetition rate reflects on the fact that the relative sideband power in the mid-IR comb strongly dominates over the near-IR one.  Fig.~\ref{fig:evencomb}i) shows the full extent of the values of the pump detunings that we have explored numerically. Within this and other sets of data and before the signal-idler regime kicks in, see detunings greater than $230$~MHz Fig.~\ref{fig:evencomb}i), we have also observed staggered Turing pattern combs with different line spacings (two, six, etc. FSRs). However, experimentally, we only captured the one with four. One possible reason for this selectivity could be an avoided crossing with another mode family, which preferentially triggers 
a certain pair of sidebands and is not accounted for in our theory.

\begin{figure*}[h]
	\centering{	\includegraphics[width=0.9\textwidth]{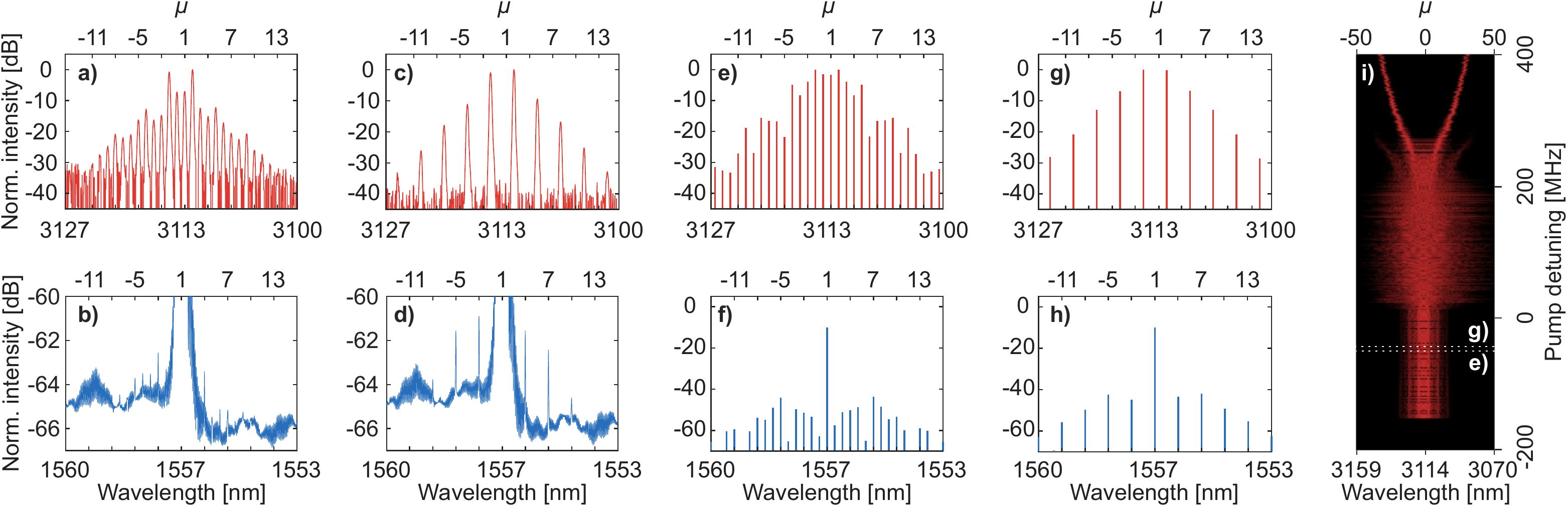}}
	\caption{\textbf{Ordinary and asymmetrically staggered frequency combs for the pump coupled to the odd mode number.} Frequency comb spectra around the pump (blue) and sub-harmonic signal (red) frequencies. In the experiment, the comb state changes from  the one FSR comb line spacing in a,b) to  the three FSR spacing in c,d) as the pump frequency is reduced across the resonance.  e-i) Numerically simulated spectra  for different  pump detunings. 
	e-f) and g-h) show the spectra for the $- 50$ and $- 44$~MHz detunings, respectively. The background in i) is at -70~dB and  $\ep_0=-0.75$~GHz. c,d) and g,h) show the asymmetric staggering.}
	\label{fig:oddcomb}
\end{figure*}

We have also investigated the case of pumping the resonator mode with an odd number, $2M+1$, by shifting the pump frequency by 1~FSR, see SM. 
In this case, the degenerate conversion is not possible, as well as the described above staggering of the two combs, since $(2M+1)/2$ is not  integer.  However, the spectral transformations induced by the pump detuning scan  looked similar to the even mode number case, cf.  Fig.~\ref{fig:oddcomb} and Fig.~\ref{fig:evencomb}. One difference was that instead of the perfect staggering and the even order Turing patterns we have now seen the Turing combs with an odd number of FSRs between the lines corresponding to the  asymmetric staggering, see Figs. \ref{fig:oddcomb} c),d) and g),h), and compare the illustrations in the middle and right columns in Fig.~\ref{f0}.

Overall, this work demonstrates a new type of the microresonator-based source for mid-IR frequency combs and the existence of the staggered two-colour frequency combs. The use of a telecom wavelength laser as a pump source might enable the realization of the setup fully integrated on a chip, with a simple single material architecture. 
The comb center wavelength, defined by the degenerate OPO phase-matching conditions, can be widely tuned by changing the resonator size. It is possible, for example, to generate a comb up to 4.3~\textmu m with a resonator size of $R=0.25$~mm. Furthermore, theory and modelling  suggest that a variety of the two-colour parametric solitons in microresonators can be expected in future experimental studies\cite{Villois2019,Smirnov2021,Dmitry2021}, in addition to the already observed ones\cite{Bruch2021}. 
One property of our resonator that could facilitate the soliton formation is that, the point of zero walk-off, i.e.,
$D_{1p}=D_{1s}$,
is achieved at 2.16~µm and 4.32~µm wavelengths for the pump and half-harmonic respectively. Furthermore, dispersion is also reduced for these wavelengths, and the zero dispersion point  is found around 5~µm. Precise control of these parameters will unlock the full potential of microcomb generation based on the $\chi^{(2)}$ nonlinearity. Advances in this and other directions will be facilitated by the   predictive capabilities of the coupled-mode theory for $\chi^{(2)}$ microresonators, see, e.g.,~\cite{Dmitry2021_theory} and references therein.


\clearpage
\section*{Supplementary information}

\newcommand{\mmu}{{\text{-}\mu}}
\newcommand{\vg}{\textsl{g}}
\newcommand{\vD}{\varDelta}
\newcommand{\nn}{\nonumber}
\newcommand{\ta}{\theta}
\newcommand{\Ta}{\Theta}
\newcommand{\sq}[1]{\sqrt{\smash[b]{#1}}}
\newcommand{\al}{\alpha}
\newcommand{\wt}{\widetilde}
\newcommand{\wh}{\widehat}
\newcommand{\cI}{{\cal I}}
\newcommand{\cN}{{\cal N}}
\newcommand{\cP}{{\cal P}}
\newcommand{\cH}{{\cal H}}
\newcommand{\cF}{{\cal F}}
\newcommand{\cE}{{\cal E}}
\newcommand{\cW}{{\cal W}}
\newcommand{\textred}[1]{\textcolor{red}{#1}}
\newcommand{\textblue}[1]{\textcolor{blue}{#1}}
\newcommand{\be}{\begin{equation}}                                       
	\newcommand{\eeq}{\end{equation}}
\newcommand{\ba}{\begin{eqnarray}}
	\newcommand{\ea}{\end{eqnarray}}
\newcommand{\bref}[1]{(\ref{#1})}

\newcommand{\bi}[1]{\bibitem{#1}}\newcommand{\lab}[1]{\label{#1}}

\newcommand{\bsub}{\begin{linenomath}\begin{subequations}}                      
		\newcommand{\esub}{\end{subequations}\end{linenomath}}     

\subsection*{Setup}

 The CdSiP$_2$ resonator, shown in Fig. 2 in the main text, was manufactured with a major radius $R = 560$~\textmu m. The raw material was grown and provided by BAE Systems and the manufacturing process is detailed in Ref. \cite{Amiune2021}. We determined a quality factor $Q_\text{p} = 6 \times 10^6$ and a free spectral range (FSR) of 27.0~GHz for 1550~nm wavelength and extraordinary polarization by scanning the pump laser over 30~GHz.

The experimental setup is sketched in Fig. 2 in the main text. In order to achieve optical parametric oscillation, we rely on birrefringent phase-matching. We pump the resonator with a continuous wave 1550~nm laser (Toptica DL Pro) with extraordinary polarization and couple light in with a silicon prism by frustrated total internal reflection. The near-IR and mid-IR beams are separated with a dielectric mirror. Ordinarily polarized generated light in the mid-infrared is detected with an optical spectrum analyzer (Yokogawa AQ6376). The OPO output wavelengths are controlled by tuning the temperature of the resonator in order to achieve operation at degeneracy. The frequency comb spectra are recorded simultaneously with two optical spectrum analyzers (Yokogawa AQ6370D and AQ6376).

\subsection*{Model}
We follow here the model, which detailed derivation from Maxwell equations can be found in \cite{Dmitry2021_theory}, and which has been previously studied in, e.g.,~\cite{Amiune2021,opo}.
We assume that $\om_p$ is the  pump laser frequency and 
$\om_{0p}$ is the frequency of the resonator mode with the number $2M$. $2M$ equals the number of wavelengths fitting along the  ring circumference. We express the multimode intra-resonator electric fields of the pump, $B$,  and half-harmonic signal, $A$, as 
\begin{equation}
\begin{split}
&e^{iM\vta-i\frac{1}{2}\om_p t}A(\vta,t)+c.c.=e^{iM\vta-i\frac{1}{2}\om_p t}\sum_\mu a_\mu(t) e^{i\mu\vta}+c.c.,\\ 
&e^{i2M\vta-i\om_p t}B(\vta,t)+c.c.=e^{i2M\vta-i\om_p t}\sum_\mu b_\mu(t) e^{i\mu\vta}+c.c..
\end{split}
\lab{field}
\end{equation}
Here, $\vta=(0,2\pi]$ is the angular coordinate and 
$\mu=0,\pm 1,\pm 2,\dots$ is the relative mode number.

$\om_{s\mu}$ and $\om_{p\mu}$ are the respective signal and pump spectra
of the resonator frequencies,
\begin{equation}
\om_{\mu s}=\om_{0s}+\mu D_{1s}+\tfrac{1}{2}\mu^2 D_{2s},~~
\om_{\mu p}=\om_{0p}+\mu D_{1p}+\tfrac{1}{2}\mu^2 D_{2p},
\lab{om}\end{equation}
where,  $D_{1s,1p}/2\pi$ are the free spectral ranges, FSRs, and $D_{2s,2p}$ are dispersions. 

Coupled-mode equations governing the evolution of $a_\mu(t)$, $b_\mu(t)$ are~\cite{Dmitry2021_theory}
\begin{equation}
\begin{aligned}
	i\p_t a_{\mu}=&(\om_{\mu s}-\tfrac{1}{2}\om_p)a_{\mu} - \frac{i\kappa_s}{2} a_{\mu} -\gamma_s\sum_{\mu_1 \mu_2}\wh\delta_{\mu,\mu_1-\mu_2}b_{\mu_1}a^*_{\mu_2},\\
	i\p_t b_{\mu}=&(\om_{\mu p}-\om_p)b_{\mu} - \frac{i\kappa_b}{2} \big(b_{\mu}-\wh\delta_{\mu,\mu'}\cH_{\mu'}\big) \\
	&-\gamma_p\sum_{\mu_1 \mu_2}\wh\delta_{\mu,\mu_1+\mu_2}a_{\mu_1}a_{\mu_2},
\end{aligned}
\lab{tp1}
\end{equation}

where $\wh\delta_{\mu,\mu'}=1$  for $\mu=\mu'$ and is zero otherwise.
$\cH_{\mu'}$ is the pump parameter,
$\cH^2_{\mu'}=\cF_p W/2\pi$, where
$W$ is the laser power, and $\cF_p=D_{1p}/\kappa_{p}$ is the pump finesse~\cite{Dmitry2021_theory}.  Pumping to the even mode number, $2M$, corresponds to $\mu'=0$ and pumping to the odd mode, $2M+1$, 
corresponds to $\mu'=1$.

The frequency matching, i.e. phase matching, parameter for the non-degenerate parametric process initiated by, e.g., the $\mu=0$ mode in the pump field is defined as
\begin{equation}
\ep_\mu=\om_{\mu s}+\om_{-\mu s}-\om_{0p}
=\frac{c}{R}\left[
\frac{M+\mu}{n_{M+\mu}}+
\frac{M-\mu}{n_{M-\mu}}
-\frac{2M}{n_{2M}}
\right].
\label{ep}
\end{equation}
Here, $n_m$  is the effective refractive index taken for the frequencies of the  modes with the absolute numbers $m=M\pm\mu$ (signal and idler) and $2M$ (pump), $c$ is the vacuum speed of light and $R$ is the resonator radius. 
Arranging for
\begin{equation}
\ep_0=2\om_{0s}-\om_{0p}=\frac{c}{R}\left[
\frac{2M}{n_{M}}
-\frac{2M}{n_{2M}}
\right]=0,\text{~i.e.,~} n_M=n_{2M},
\end{equation}
corresponds to the exact matching for the degenerate parametric conversion. We shall note, that Eq. (1) in the main text has been
derived in \cite{opo} for the pumping to the even mode only, $\mu'=0$. 

Fourier transforming the coupled mode equations, Eq.~(\ref{tp1}), to the physical space one arrives to the following set of partial-differential equations~\cite{Dmitry2021_theory},
\begin{equation}\label{eq:dynamics}
\begin{aligned}
i\p_tA =& \left(\om_{0s}-
\tfrac{1}{2}\om_p-iD_{1s}\p_\vta
-\frac{1}{2}D_{2s}\p_{\vta}^2\right)A\\
&-\frac{i\kappa_s}{2}A-\gamma_sBA^*~, \\
i\p_t B=&\left(\om_{0p}-\om_p-iD_{1p}\p_{\vta}-
\frac{1}{2}D_{2p}\p_{\vta}^2\right)B\\
&- \frac{i\kappa_p}{2}
\left(B-\cH_{\mu'} e^{i\mu'\vta}\right) -\gamma_p A^2.
\end{aligned}
\end{equation}

The calculation of the resonator's FSR, dispersion and phase-matching temperatures are carried out by approximating the resonance frequencies as in \cite{WGRmodes}, including the Sellmeier equation\cite{CSPsellmeier2018} and thermal expansion coefficients\cite{Zawilski2010} for CdSiP$_2$ and assuming the fundamental transverse mode numbers for the pump and signal waves.
The transmittance in Fig.~1d is calculated as the power dissipated over the roundtrip,
\begin{equation}
T= 1-\frac{2\pi\eta}{W}
\sum_{\mu}\left(
\frac{\left|b_{\mu}\right|^2}{\mathcal{F}_p}+
\frac{\left|a_{\mu}\right|^2}{\mathcal{F}_s}\right),
\end{equation}
where $\eta = 0.3$ is the fitting parameter.

\begin{table}[h!]
    \caption{ Resonator parameters.}
\centering
    \begin{tabular}{|m{2cm}|m{3.5cm}|m{3.5cm}|} 
    \hline
     & Pump (1556~nm) & Signal (3112~nm) \\
    \hline
    Repetition rate & $D_{1p}/2\pi=$27.06~GHz & 
    $D_{1s}/2\pi=$27.57~GHz\\
    Dispersion & $D_{2p}/2\pi=-$378~kHz & 
    $D_{2s}/2\pi=-$164~kHz\\
    Linewidth & 
    $\kappa_p/2\pi=$55~MHz & 
    $\kappa_s/2\pi=$64~MHz\\
    Nonlinearity & 
    $\gamma_p/2\pi=$1~GHz $\text{W}^{-1/2}$ & 
    $\gamma_s/2\pi=$2~GHz $\text{W}^{-1/2}$\\
    \hline
    \end{tabular}
    \label{table}
\end{table}

\bibliography{CSP_comb_ref}

\end{document}